\documentclass{aa}
\usepackage{times}
\usepackage{graphics}
\usepackage{xspace}
\usepackage{epsfig}
\usepackage{jwaabib}
\usepackage{rotating}
\usepackage{dcolumn}

\newcommand{\sO}{$\sigma$\,Ori}
\newcommand{\rOp}{$\rho$\,Oph}

\begin{document}

\title{X-ray Emission near the Substellar Limit: \\ 
The $\sigma$ Orionis and Taurus Star Forming Regions}

\author{F. Mokler \and B. Stelzer}

\institute{Max-Planck-Institut f\"ur extraterrestrische Physik,
  Postfach 1312,
  D-85741 Garching,
  Germany}

\offprints{F. Mokler}
\mail{F. Mokler, fmokler@mpe.mpg.de}
\titlerunning{X-ray emission from young, very low-mass objects}

\date{Received $<$19-04-02$>$ / Accepted $<$11-06-02$>$}

\abstract{
We have carried out an extensive search for X-ray emission from young, very 
low-mass objects near and beyond the substellar limit, making use of archived 
{\em ROSAT} PSPC and HRI observations pointed at Brown Dwarfs and Brown Dwarf 
candidates in the young $\sigma$\,Orionis and Taurus-Auriga associations. In 
\sO \ we identify three Brown Dwarf candidates with X-ray sources; in 
Taurus-Auriga we add one further X-ray detection of a Brown Dwarf to the list 
published earlier. We combine this data with all previously X-ray detected 
Brown Dwarfs and Brown Dwarf candidates in young stellar associations and star 
forming regions to perform a study of stellar activity parameters on the as yet
 largest sample of young, very low mass objects. A similar relation between 
X-ray and bolometric luminosity, and H${\alpha}$ emission, respectively, as is 
known for T Tauri stars seems to hold for young objects down to the substellar 
limit, too. 
No signs for a change in X-ray activity are found on the transition to 
substellar masses. 
\keywords{X-rays: stars -- stars: formation, low-mass, Brown Dwarfs, coronae, activity}
}

\maketitle

\section{Introduction}\label{sect:intro}

Late-type stars exhibit strong signs of magnetic activity such as
H$\alpha$, Ca\,II or X-ray emission from hot thermal plasma
confined in magnetic fields on the star. For fully convective stars (with 
spectral type $\sim$\,M3 and later) the change in interior structure is 
expected to result in a change of the field sustaining dynamo and, therefore,
of the emission properties. However, no clear change in common activity 
diagnostics is found at these spectral types. \citey{Gizis00.1}, 
\citey{Basri01.1} and \citey{Mohanty02.1} have observed a decline in 
H$\alpha$ emission for old ($>1$\,Gyr) L-type objects in the field, 
suggesting that a decline in dynamo activity sets in beyond the boundary 
where the objects become fully convective and close to the borderline to 
substellar masses. An X-ray study of K- and early M-stars within 
the solar neighborhood by \citey{Fleming95.1} did not unveil any change in 
X-ray activity for objects on the transition towards fully convective energy 
transport. It is unclear to date how activity in fully convective 
and substellar objects depends on parameters such as rotation or age. 
Studying these relations is essential to an understanding of the underlying
dynamo mechanism.

An important measure for magnetic activity on late-type stars of all
ages is X-ray emission which is usually explained by emission from a hot plasma.
Within the last few years, X-ray emission has also been detected from
Brown Dwarfs (BD) and BD Candidates  
(see e.g. Neuh\"auser \& Comer\'on 1998,
Neuh\"auser et al 1999). In these early studies with {\em ROSAT} 
only BDs in the Chamaeleon, Taurus and $\rho$\,Ophiuchus star 
forming regions were detected, while all older substellar objects in stellar 
associations such as the Pleiades and in the field are X-ray quiet down to the
{\em ROSAT} detection limit. More recent observations with {\em Chandra}
revealed X-ray emission from further young BDs 
in Orion (\cite{Garmire00.1}, \cite{Feigelson02.1}), 
$\rho$\,Oph (\cite{Imanishi01.1}), and IC\,348 (\cite{Preibisch01.1}; 
\cite{Preibisch02.1}).
To date only one old field BD is known to emit X-rays, namely LP\,944-20 which was detected only during a flare (\cite{Rutledge00.1}).
Here, we extend the investigations with a study
of X-ray emission from the very low-mass (VLM) members
in the $\sigma$\,Orionis association, 
and an update of the X-ray emitting BD population in Taurus. 

The $\sigma$\,Ori star forming region was discovered by 
\citey{Walter97.1}, and was shown to be rich in 
X-ray sources. 
Photometric and spectroscopic observations by \citey{Bejar99.1}, \citey{Zapatero00.1},
and \citey{Bejar01.1} have revealed $\sim$\,80 objects in the
low mass regime close to and below the hydrogen burning mass limit (HBML) 
in the OB1b association near the multiple star $\sigma$\,Orionis at a mean distance modulus of $DM = 7.73$ measured by {\em HIPPARCOS}, corresponding to $352$\,pc (\cite{Bejar99.1}).
The age of this cluster is estimated to be $1-7$\,Myrs
where the upper age limit is given by constraints of the
central star $\sigma$\,Orionis: being of spectral type O9.5
and still in the hydrogen burning phase, it cannot be older than $5-7$\,Myrs
(see \cite{Bejar01.1}, \cite{Barrado01.1}, and references therein).
According to \citey{Bejar01.1},
the position of all members in the H-R diagram is best
compatible with an isochrone for 5\,Myrs. Based on measurements 
of lithium abundances in \sO \ members
\citey{Zapatero02.1} give as most likely age 
$2-4$\,Myrs with an upper limit of 8\,Myrs.
At this young age, even VLM objects are still very luminous. 
In addition, the $\sigma$\,Ori association is hosted in a region of very
low extinction ($E_{B-V}=0.05$; \cite{Lee68.1}). This provides excellent 
conditions
to study activity in faint VLM
objects despite the considerable distance to the cluster.

Taurus-Auriga is one of the nearest ($d$ = 140\,pc; \cite{Elias78.1}) and 
most-studied regions of star formation. 
The X-ray emission from late-type pre-main sequence stars in Taurus-Auriga 
was recently discussed by \citey{Stelzer01.1} (hereafter SN01).
SN01, whilst concentrating on the G- to early M-type T Tauri Stars, 
put forth the {\em ROSAT} PSPC detection of several objects 
with spectral types beyond M5, i.e. near the substellar limit. 
Four additional BDs have been discovered in that region (\cite{Martin01.2}) since.
In this paper we present a detailed analysis of the X-ray activity 
of all VLM objects in Taurus-Auriga including the new BDs.
In the following we used the term 'very-low mass' for objects 
with spectral type later than M4.

In Sect.~\ref{sect:sample}, we outline the criteria we used when selecting our data samples.  
In Sect.~\ref{sect:data}, we describe the {\em ROSAT} data analysis.
We present the results for both $\sigma$\,Orionis and Taurus 
in Sect.~\ref{sect:results}. 
In Sect.~\ref{sect:discussion} we perform a comparative study of 
the X-ray properties of all young BDs and BD candidates detected so far, including
tests for variability and an investigation of correlations
with other activity parameters.
A summary of our results is given in Sect.~\ref{sect:conclusions}.

\section{Sample Selection}\label{sect:sample}

As we searched for X-ray emission from VLM objects in the 
$\sigma$\,Orionis and Taurus star forming regions our sample consists of all objects with spectral type $\sim$\,M5 and later. 
At an age of 5\,Myrs all objects with spectral type M7 and later 
can be considered to be substellar (\cite{Basri00.1} and references therein). Therefore, our
sample includes the transition from VLM stars to BDs.

For about one third of the VLM $\sigma$\,Ori members the spectral 
types are known, and cover the range from M4.5 to L6 
(\cite{Bejar99.1}, \cite{Bejar01.1}, \cite{Martin01.1} and \cite{Barrado01.1}).
A large fraction of the cluster members, however, has been observed only
photometrically, so far.
For the X-ray detected objects we estimated the spectral types by comparing their $I-J$, $I-K$ and $R-I$ colors with
those of the objects in \sO \ with known spectral type, assuming negligible extinction.

The Taurus sample consists of all objects beyond spectral type M4
included in SN01, and the newly identified BDs from \citey{Martin01.2}.


\section{Observations and Data Analysis}\label{sect:data}

We systematically searched the {\em ROSAT} archive for deep observations
 including VLM objects in Tau and \sO . 
The majority of {\em ROSAT} PSPC observations with BDs or BD candidates in Taurus can be found in SN01. 
We added to this list all PSPC pointings including any of the objects from 
\citey{Martin01.2}, i.e. {\em ROSAT} Observation IDs 
201016p and 201017p. Besides we examined all HRI pointings 
with Taurus objects from our sample (202156h, 202031h, 201090h, 201046h, 
201623h-1/-2, 201089h, 201617h/-1).

Eight PSPC pointings are available for the $\sigma$\,Orionis region:
180023p-0/-1, 200198p, 200932p,
201151p-0/-1, 900198p, and 900386p.
Careful inspection showed 
that only pointings 201151p-0 and 201151p-1 are
useful for aiming at faint detections.
In the remaining cases, the objects are
either located at a large off-axis angle ($\Delta \ge 39^{'}$)
within the field of view (FOV), or the net exposure time at the position
of the objects is too low as it was covered by the ribs of the telescope 
over a substantial fraction of the exposure time. 
Hence, we exclusively used pointings 201151p-0 and 201151p-1  
from this list. In order to improve the sensitivity both pointings were 
merged to add up to one pointing with total exposure time
of $\sim 33$\,ksec. In the following we refer to this combined observation as
201151p.
All available 35 HRI pointings in $\sigma$\,Ori are centered at
the same sky position as 201151p.
They were all added up to yield a pointing with total exposure time of
$\sim 81$\,ksec. This pointing will be named after the last one
of the whole sequence, 201915h.

The data were analysed in the {\em MIDAS/EXSAS} environment. 
We used a combination of the {\em map}, {\em local} and 
{\em maximum likelihood} 
detection algorithm described by \citey{Cruddace87.1}. 
The source detection threshold was set to $ML \ge 7.4$ (PSPC) and $ML \ge 5.0$ (HRI), respectively (corresponding to a reliability $> 99 \%$; see \cite{Neuhaeuser95.1}). 
Source detection was
performed in the broad (0.1 to 2.0\,keV), soft (0.1 to 0.4\,keV), hard~1 
(0.5 to 0.9\,keV) and hard~2 (0.9 to 2.0\,keV) energy bands of the PSPC, and
in the broad band (0.1 to 2.0 keV) of the HRI.
In order to get qualitative information on the X-ray spectrum, 
we calculated the PSPC hardness ratios, $HR\,1$ and $HR\,2$,
 from the X-ray count rates in the soft, hard~1 and hard~2 band:
\begin{equation}
HR\,1=\frac{Z_{h1}+Z_{h2}-Z_{s}}{Z_{h1}+Z_{h2}+Z_{s}}
\end{equation}
and
\begin{equation}
HR\,2=\frac{Z_{h2}-Z_{h1}}{Z_{h2}+Z_{h1}}
\end{equation}
where $Z_{\rm x}$ denotes the count rate in the corresponding energy band.   

For the identification of a BD or BD candidate with an X-ray source
we allowed for a maximum offset of $30^{''}$ in the PSPC pointing
and $10^{''}$ in the HRI pointing,
for sources in the inner area of the FOV, i.e. at off-axis angles 
smaller than $30^{\prime}$ (PSPC) and $6^{\prime}$ (HRI),
corresponding to the spatial resolution of the respective detectors.
For X-ray sources at larger off-axis angles we gradually increased the 
identification radius as described by SN01 
to take account of the degradation of the PSPC point spread function (PSF).
The HRI data were treated in an analogous way.
Visual inspection of the X-ray images showed that the reliability of 
automatic source detection is limited for faint X-ray sources in the immediate 
neighborhood of a bright one.  
Therefore, we performed an additional check of the {\em ROSAT} images by eye.
In the case of S\,Ori\,\,03, a faint X-ray source can be clearly discerned in the 
PSPC pointing. However, it was not found by the 
detection procedure. For this source we extracted the photons 
from a circular region centered on the optical position of S\,Ori\,03 
(see Table~\ref{tab:bds}).

For the photon extraction radius, we used the 99\,\% 
 quantile radius of the point spread function at 1\,keV.

When converting the X-ray count rates into fluxes, 
we assumed a Raymond-Smith spectrum representing a thermal plasma at 1\,keV.
In contrast to the analysis of SN01 we did not attempt to split the 
X-ray counts among the components in binary systems 
as the multiplicity of most objects in our sample is unknown.

\section{X-ray Identifications in $\sigma$\,Orionis and Taurus}\label{sect:results}

Following the procedure described in the previous section 
we detected X-ray sources near seven VLM \sO \ members 
and 13 BD candidates and BDs in Taurus. 
We double-checked with the {\em SIMBAD} and {\em GSC} catalogues for other
possible counterparts to these X-ray sources within a radius of $30''$. 
In ambiguous cases, we also consulted the {\em Digitized Sky Survey} ({\em DSS}).
Below we describe the results for the individual objects.

\subsection{$\sigma$\,Orionis}\label{subsect:sori}

With Table~\ref{tab:sori_all} we provide a list of all VLM $\sigma$\,Ori
members inside the error box of a {\em ROSAT} detection.
\begin{table*}
\begin{center}
\begin{tabular}{lrrlrlr}\hline
Designation         & $\alpha_{\rm x(2000)}$ & $\delta_{\rm x(2000)}$ & ROR &   ${\rm Offset_{x}}$ & GSC,Simbad,DSS & $ {\rm Offset_{cat}}$ \\ 
    &    &   &    & [$^{\prime\prime}$]  & & [$^{\prime\prime}$] \\ \hline
S\,Ori\,03   & 05 39 20.8 & -02 30 35 & 201151p & 0.0 & $-$&$-$ \\ 
$\prime\prime$      & 05 39 20.4 & -02 30 36 & 201915h & 5.6 & $-$&$-$ \\        
S\,Ori\,08   & 05 39 07.6 & -02 28 24  &201151p & 23.9 & DSS&$\sim 1.5$ \\               
S\,Ori\,51   & 05 39 02.7 & -02 29 55 &201151p  & 26.1 & DSS&$\sim 2.0$ \\             
S\,Ori\,07     & 05 39 07.6 & -02 32 37 &201915h  & 9.9 & Haro 5-19 & 2.3\\         
S\,Ori\,43   & 05 38 14.0 & -02 35 07 &201151p  & 4.1 & 12906026 (GSC)& 2.2\\        
$\prime\prime$   & 05 38 13.7 & -02 35 08 &201915h  & 3.9 & $\prime\prime$ & $\prime\prime$ \\
S\,Ori\,053926.8-022614$^{*}$ & 05 39 25.7 & -02 26 12 & 201151p & 17.2 & $-$&$-$ \\               
S\,Ori\,053948.1-022914$^{*}$ & 05 39 48.1 & -02 29 07 & 201151p & 7.1 & $-$&$-$ \\               
\hline
\end{tabular}
\caption{VLM $\sigma$\,Ori members in the error box of a {\em ROSAT} source, and further potential optical counterparts from the {\em SIMBAD}, {\em GSC} or {\em DSS} catalogues, respectively. We give object name, X-ray position, {\em ROSAT} observation request number, offset between X-ray and optical position of the object in \sO \ , optical counterpart from catalogue, and offset between X-ray source and catalogued counterpart. $^{*}$ Note that we use the abbreviations $\sigma$\,Ori\,053926.8 and $\sigma$\,Ori\,053948.1 in the text.}
\label{tab:sori_all}
\end{center}
\end{table*}
In the cases of S\,Ori\,07 and S\,Ori\,43, we found catalogued
objects whose optical
positions are closer to the X-ray position than that of the \sO \ members.
The {\em GSC} and {\em SIMBAD} object closest to S\,Ori\,07 is 
Haro 5-19, an emission line star, 
with an offset of $2.3^{\prime\prime}$ from the X-ray position, 
whereas the offset of S\,Ori\,07 is $9.9^{\prime\prime}$. Hence, the X-ray emission probably 
originates from the Haro star.
The closest counterpart to the X-ray source near S\,Ori\,43 is the {\em GSC} object 12906026 with an offset of $2.2^{\prime\prime}$, whereas the offset between S\,Ori\,43 and the X-ray source is $4.1^{\prime\prime}$ (PSPC) and $3.9^{\prime\prime}$ (HRI), respectively.  
Checking the {\em DSS} image we conclude from the
relative position of the X-ray source, the {\em GSC} object 
and S\,Ori\,43 that the X-ray source is more likely to be assigned to the 
{\em GSC} source than to S\,Ori\,43.
For the remaining five X-ray identified \sO \ members, no other counterparts 
neither in {\em SIMBAD} nor in the {\em GSC} were found. However, the 
offsets of S\,Ori\,08 and S\,Ori\,51 with respect to the X-ray
position are near the limit of the {\em ROSAT} error box. 
At the position of the X-ray source 
near S\,Ori\,51 an unknown optical object can 
clearly be distinguished on the {\em DSS} image. 
For S\,Ori\,08, there is an optical source discernable both at its optical 
position and at the position of the nearby X-ray source. 
Therefore, we assign the X-ray emission to the {\em DSS} object at the same position in both cases. 
The identification of S\,Ori\,03 with the X-ray source is unequivocal from the 
inspection of the {\em DSS} images. 
In the cases of S\,Ori\,J053926.8 and  S\,Ori\,J053948.1 no other clear counterpart was found in the {\em DSS} either.   
In the following we discuss only these three unambiguous detections. For the 
other cases, where the nature of the additional optical counterparts is not 
known, we plan follow-up spectroscopy. 

In Table~\ref{tab:bds} we list the X-ray properties of the three 
{\em ROSAT} detections which were clearly 
identified with $\sigma$\,Ori members.
All of them were detected in the PSPC pointing,
and one object, S\,Ori\,03, was also detected in the HRI pointing.
In Fig.~\ref{fig:sO_PSPC}
we display the portion of the PSPC and HRI images that include these 
$\sigma$\,Ori detections.

\begin{sidewaystable*}\scriptsize
\begin{center}
\caption{X-ray parameters for all {\em ROSAT} detected VLM \sO \ members and Taurus BD candidates and BDs, sorted by right ascention. The columns contain object name, spectral type, {\em ROSAT} observation request number, X-ray position, offset between X-ray position and optical counterpart, offaxis position of the objects within the corresponding FOV, maximum likelihood of existence, exposure time, X-ray counts, X-ray-luminosity, $\lg{(\frac{L_{\rm x}}{L_{\rm bol}})}$, hardness ratios $HR$\,1 and $HR$\,2, and the results from the KS-test. The error in the X-ray luminosity does not include the uncertainty in the distance and is discussed in the text.}
\label{tab:bds} 
\begin{tabular}{llrrrrrrrrcrrrr}
\hline
Designation & Sp.type & \multicolumn{1}{c}{{\em ROSAT}} & \multicolumn{2}{c}{X-ray Position}                                                 & Offset & Offax & ML & Expo & Cts & \multicolumn{1}{c}{$\lg{L_{\rm x}}$} & \multicolumn{1}{c}{$\lg{(\frac{L_{\rm x}}{L_{\rm bol}})}$} & \multicolumn{1}{c}{HR\,1} & \multicolumn{1}{c}{HR\,2} & KS-result \\ 
            &         & \multicolumn{1}{c}{ROR}      & \multicolumn{1}{c}{$\alpha_{\rm 2000}$} & \multicolumn{1}{c}{$\delta_{\rm 2000}$}  & \multicolumn{1}{c}{[$^{\prime\prime}$]} & \multicolumn{1}{c}{[$^\prime$]} & & [ks] & & \multicolumn{1}{c}{[erg/s]} &                                &       &       &           \\ \hline
\multicolumn{15}{c}{\sO onis}\\ \hline
S Ori03          & M5-6$^{a}$ & 201915h & 05 39 20.4 & $-$02 30 36  & 5.6     & 10.8 &  6 & 83.4 &  38.0 & $29.37 \pm 0.16$ & $-$3.18 & $--$ & $--$ & $-$ \\
$^{\prime\prime}$& $^{\prime\prime}$   & 201151p & 05 39 20.8 & $-$02 30 35  & 0.0$^b$   & 10.9 &  59 & 33.4 &  63.5 & $29.53 \pm 0.08$ & $-$3.02 & $>0.70$ & $0.06 \pm 0.16$ &      $-$  \\
SOri J053926.8   & M5$^a$          & 201151p & 05 39 25.7 & $-$02 26 12  & 17.2    & 14.6 &  20 & 31.6 &  31.8 & $29.16 \pm 0.16$ & $-$1.95 & $>0.99$  &$0.46 \pm 0.22$ &  $-$ \\
SOri J053948.1   & M6$^a$          & 201151p & 05 39 48.1 & $-$02 29 07  & 7.1     & 17.5 &  42 & 28.9 &  65.0 & $29.54 \pm 0.09$ & $-$1.49 & $0.69 \pm 0.20$ & $0.30 \pm 0.16$ &    $-$ \\
\hline
\multicolumn{15}{c}{Taurus}\\ \hline
%
%
FNTau              & M5   & 200949p          &  04 14 14.7 & +28 27 54 &  5.4 & 15.6 & 101 &  5.4 &  52.3 & $29.36 \pm 0.15$ & $-3.97$ & $> 0.62$          & $ 0.31 \pm 0.14$ & $-$\\
RXJ0416.5+2053     & M5-6 & 201316p         &  04 16 30.2 & +20 53 07 & 38.8 &  24.0 &   9 &  3.1 &   7.3 & $28.74 \pm 0.39$ & $    *$ & $>-0.24$          & $ 0.07 \pm 0.42$ & $-$\\
$\prime \prime$     & $\prime \prime$ & 201504p         &  04 16 31.5 & +20 53 36 & 51.6 &  43.3 &  20 &  5.1 &  20.3 & $28.97 \pm 0.36$ & $    *$ & $>-0.69$          & $ 0.75 \pm 0.22$ & $-$\\
V410x-ray3         & M6.5 & 200001p-0/p-1    &  04 18 08.4 & +28 26 00 &  7.2 &  4.9 &  74 & 28.2 &  57.3 & $28.68 \pm 0.16$ & $-3.83$ & $> 0.33$          & $ 0.02 \pm 0.16$ & $>99\%$ \\
StromAnon13        & M5   & 200001p-0/p-1    &  04 18 18.1 & +28 28 41 & 12.1 &  3.2 &  13 & 30.8 &   9.3 & $27.85 \pm 0.55$ & $-4.43$ & $>-0.41$          & $ 0.81 \pm 0.39$ & $-$ \\
Kim3-89            & M5   & 200001p-0/p-1    &  04 19 01.5 & +28 19 44 &  4.1 & 10.3 &  62 & 27.9 &  54.1 & $28.78 \pm 0.10$ & $-4.19$ & $ 0.92 \pm 0.26$ & $ 0.23 \pm 0.17$ & $-$\\  
V410x-ray5a        & M5   & 200001p-0/p-1    &  04 19 01.7 & +28 22 33 &  4.1 &  8.6 & 141 & 26.3 &  83.5 & $28.87 \pm 0.15$ & $-3.52$ & $> 0.46$          & $ 0.35 \pm 0.12$ & $-$\\
$\prime \prime$        & $\prime \prime$  & 202156h          &  04 19 02.0 & +28 22 34 &  0.1 &  6.6 &  14 &  7.4 &  15.2 & $29.66 \pm 0.15$ & $-2.73$ & $--$             & $--$             & $-$\\
J2-157             & M5.5 & 200442p          &  04 20 53.0 & +17 46 40 &  6.0 & 16.0 & 128 & 18.1 & 133.2 & $29.18 \pm 0.06$ & $-3.25$ & $ 0.09 \pm 0.10$ & $ 0.09 \pm 0.13$ & $-$\\   
$\prime \prime$        & $\prime \prime$ & 201370p-0/p-1    &  04 20 54.3 & +17 46 06 & 42.9 & 39.5 &  13 & 12.6 &  81.4 & $29.14 \pm 0.13$ & $-3.29$ & $ 0.16 \pm 0.25$ & $ 0.26 \pm 0.25$ & $-$\\
MHO-4     & M5   & 200443p          &  04 31 22.9 & +18 00 07 & 21.8 & 32.7 &  13 & 15.2 &  47.5 & $28.91 \pm 0.20$ & $-3.28$ & $ 0.54 \pm 0.39$ & $-0.39 \pm 0.26$ & $>99\%$ \\   
$\prime \prime$     &  $\prime \prime$  & 201313p/900353p  &  04 31 24.2 & +18 00 24 &  3.9 &  6.5 &  82 & 11.5 &  39.9 & $29.04 \pm 0.09$ & $-3.14$ & $> 0.66$          & $ 0.09 \pm 0.17$ & $-$\\
V927Tau            & M5.5 & 200694p-0/p-1    &  04 31 23.6 & +24 10 56 &  2.8 & 21.7 &  62 &  2.6 &  31.5 & $29.51 \pm 0.11$ & $-3.77$ & $ 0.61 \pm 0.19$ & $-0.19 \pm 0.20$ & $>98\%$ \\
MHO-5              & M6   & 201313p/900353p  &  04 32 15.1 & +18 12 48 & 13.3 & 16.6 &  14 &  9.9 &  12.2 & $28.60 \pm 0.23$ & $-3.47$ & $>-0.30$          & $-0.08 \pm 0.34$ & $-$\\
RXJ0432.7+1809     & M5   & 201313p/900353p  &  04 32 40.6 & +18 09 23 &  5.0 & 17.6 &  26 &  7.1 &  17.7 & $28.89 \pm 0.20$ & $    *$ & $ 0.96 \pm 0.49$ & $0.66$           & 100\% \\
LH0429+17          & M9   & 200443p          &  04 32 50.4 & +17 30 10 & 10.1 &  4.9 &   8 & 20.2 &  18.7 & $28.18 \pm 0.26$ & $-2.16$ & $-0.24 \pm 0.35$ & $ 0.58 \pm 0.57$ & $-$\\    
CFHT-BD-Tau~4     & M7   & 201016p          &  04 39 47.2 & +26 01 49 & 10.0 &  6.8 &  54 & 10.0 &  28.1 & $28.91 \pm 0.12$ & $-3.33$ & $ 0.74 \pm 0.26$ & $ 0.45 \pm 0.21$ & $-$\\

\hline
\multicolumn{15}{l}{$^a$ Spectral type estimated from colors $I-J$, $I-K$, and $R-I$, respectively.}\\
\multicolumn{15}{l}{$^b$ No offset between optical and X-ray position as X-ray counts were extracted at the optical position (see text).}\\
\end{tabular}
\end{center}
\end{sidewaystable*}

\begin{figure*}
\parbox{19.0cm}{
\parbox{6.0cm}{
a
}
\parbox{6.0cm}{
b
}
\parbox{6.0cm}{
c
}}
\parbox{19.0cm}{
\parbox{6.0cm}{
d}
\parbox{11.5cm}
{\caption{Partial view of {\em ROSAT} PSPC (broad band) and HRI images with X-ray detected BDs and BD candidates from the \sO \ and Tau star forming regions.The optical positions of the VLM objects are marked by crosses. The circles mark the 99\% 
quantile area of the associated X-ray source.
(a) PSPC pointing 201151p including the three X-ray detected BD candidates in \sO ; radii of X-ray sources: S\,Ori\,03 $\sim 38^{\prime \prime}$, S\,Ori\,J053926.8 $\sim 50^{\prime \prime}$ and S\,Ori\,J053948.1 $\sim 62^{\prime \prime}$.
(b) HRI detection of S\,Ori\,03 in 201915h; radius of X-ray source $\sim 22^{\prime \prime}$ 
(c) PSPC pointing 201016p with CFHT-BD-Tau~4; radius of X-ray source $\sim 32^{\prime \prime}$
(d) PSPC pointing 200443p with LH0429+17: radius of X-ray source $\sim 31^{\prime \prime}$.
}
\label{fig:sO_PSPC}
}
}
\end{figure*}

\subsection{Taurus}\label{subsect:taurus}

The study by SN01 showed that about half of the BD candidates in 
the Taurus region known at that time were detected with the 
PSPC in pointed observations. 
Their X-ray luminosities are in agreement 
with the earlier study by \citey{Neuhaeuser99.1}.
We adopted the values listed in SN01, and added an analysis of HRI 
observations of the Taurus sample as described in Sect.~\ref{sect:sample}.
This analysis resulted in one additional HRI-detection of the M5-type object, V410x-ray5a,
which had already been detected during a PSPC observation.

Furthermore, we searched all PSPC and HRI observations 
for X-ray emission 
from the position of the BDs newly discovered by \citey{Martin01.2},
and found that one of them, CFHT-BD-Tau~4, is detected during a $\sim 10$\,ksec PSPC pointing. 
Cross-correlation with the {\em GSC} and {\em SIMBAD} as well as visual inspection 
of the corresponding {\em DSS} images showed that in all cases the X-ray emitter
can clearly be identified with the Taurus object.
In Fig.~\ref{fig:sO_PSPC} we show the portion of the PSPC image 
with the X-ray detection of CFHT-BD-Tau~4. 
In Table~\ref{tab:bds} we summarize the X-ray parameters of all 
BDs and BD candidates in Taurus-Auriga, which were detected in at least one 
PSPC or HRI pointing, i.e. both the objects from SN01 and the new
detections.

\section{Discussion}\label{sect:discussion}

We compose the largest currently available sample 
of X-ray emitters near and below the substellar limit
by combining all X-ray detections of VLM (spectral type M5 and 
beyond) objects  in star forming regions and young stellar associations, all with ages between 1 and 10\,Myrs. 
In addition to the objects in $\sigma$\,Ori and Taurus presented in this paper, X-ray emission has been reported from BDs and BD candidates in Cha\,I, the Orion Nebular Cluster (ONC), 
$\rho$\,Oph, and IC\,348 observed with {\em ROSAT} and/or {\em Chandra}.
According to \citey{Neuhaeuser98.1} and 
\citey{Comeron00.1} the Cha\,I star forming cloud hosts 
seven 
X-ray emitting BDs and BD candidates. 
\citey{Feigelson02.1} reported about
30 X-ray emitting VLM objects in the ONC detected with {\em Chandra}. 
In the \rOp \ star forming region only one BD candidate 
was found by {\em ROSAT} to emit X-rays (\cite{Neuhaeuser99.1}). 
Recent observations with {\em Chandra} by \citey{Imanishi01.1} have revealed 
more X-ray emitting BDs and BD candidates in this region. Furthermore, 
\citey{Preibisch01.1} and \citey{Preibisch02.1} have detected X-ray emission from BDs in 
IC\,348 with {\em Chandra}.

In Table~\ref{tab:all_bds_xL} we provide the mean X-ray luminosities and $\lg {\frac{L_{\rm x}}{L_{\rm bol}}}$ ratios 
for all objects with spectral type later than M4 in the different star forming
regions.
The X-ray luminosities of the \sO \ objects seem to be 
somewhat higher than in most other samples.
This could be the result of the relatively large distance
of the \sO \ association, which allows to detect
only the X-ray brightest objects with the {\em ROSAT} instruments. 
Note, that at a comparable distance in IC\,348 several BDs and BD candidates 
were detected in a $\sim$\,50\,ksec exposure with {\em Chandra}, underlining
the value of long observations at high sensitivity in the search for X-ray
emission from substellar objects. The most recent X-ray detections 
of VLM objects in the even more distant ONC were achieved with the
{\em Chandra} satellite, too.
The unusual high value of $\lg {\frac{L_{\rm x}}{L_{\rm bol}}}$ for 
the \sO \ objects, however, cannot be explained simply by selection effects and is discussed in the next section.
\begin{table}\scriptsize
\begin{center}
\caption{Mean X-ray luminosities and $L_{\rm x}/L_{\rm bol}$ values for 
the X-ray detected objects near the substellar limit in different star forming regions. 
We give also the number of detected sources ($N_{\rm D}$) and the distance.}
\label{tab:all_bds_xL}
\begin{tabular}{lrrrllr}\hline
Region & $N_{\rm D}$ & $\lg{\langle L_{\rm x} \rangle}$ & $\lg{\langle \frac{L_{\rm x}}{L_{\rm bol}} \rangle}$  & dist. & Instr. & Ref.\\
       &             & [erg/s]                          & [erg/s]                  & [pc]                    &        &       \\
\hline
Cha\,I       & 7 & 28.3 & $-$3.7  & 160  & {\em ROSAT}   & (1,2)\\
\rOp \       & 1+6  & 29.2 & $-$3.7  & 160  & {\em ROSAT}+{\em Chandra} & (3,4)  \\
IC\,348      & 7 & 28.2 & $-$3.5  & 310 & {\em Chandra} & (5)   \\
ONC          & 30 & 29.2 & $-$2.9 & 470& {\em Chandra}& (6) \\
Tau          & 13 & 29.1 & $-$3.0  & 140  & {\em ROSAT}   &      \\
\sO \        & 3 & 29.4 & $-$2.3   & 350 & {\em ROSAT}   &       \\
\hline
\multicolumn{6}{l}{(1) - \citey{Neuhaeuser98.1}, (2) - \citey{Comeron00.1},} \\ 
\multicolumn{6}{l}{(3) - \citey{Neuhaeuser99.1}, (4) - \citey{Imanishi01.1},} \\
\multicolumn{6}{l}{(5) - \citey{Preibisch01.1}, (6) - \citey{Feigelson02.1}}\\
\end{tabular}
\end{center}
\end{table}

\subsection{$L_{\rm x}-L_{\rm bol}-$Relation}\label{subsect:lb_lbol}

\begin{figure}
\begin{center}
\resizebox{8.5cm}{!}{\includegraphics{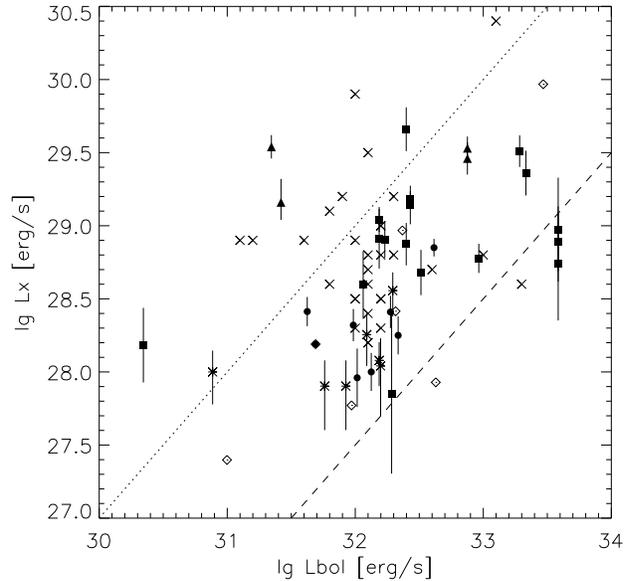}}
\caption{Correlation between $\lg {L_{\rm x}}$ and $\lg {L_{\rm bol}}$ for BDs and BD candidates observed with X-ray satellites: {\em ROSAT} observations in Taurus-Auriga (squares), Chameleon~I (filled circles), \sO \ (triangles); {\em Chandra} observations in IC348 (asterisks) and ONC (crosses). \rOp \ was observed both with {\em ROSAT} (filled diamond) and {\em Chandra} (empty diamonds). The interrupted lines mark the area typical for TTS.}
\label{fig:bol}
\end{center}
\end{figure}

In Fig.~\ref{fig:bol}, we have plotted the X-ray luminosity as a 
function of the bolometric luminosity for all presently known
X-ray detections of BDs and BD candidates in star forming regions. 
For clarity we omit the large number of non-detections. 
The dotted and dashed lines mark the range of 
$\lg{\frac{L_{\rm x}}{L_{\rm bol}}}$ from $-4.5$ to $-3$ which is typical for 
T Tauri stars (TTS). According to Fig.~\ref{fig:bol}, objects with masses down 
to the substellar limit seem to obey a very similar relation although for a 
few objects $L_{\rm x}$ is somewhat higher than expected from the
canonical relation of late-type stars.
From our sample especially three objects, namely
LH\,0429+17, SOri\,J053926.8 and SOri\,J053948.1, lie significantly 
above the saturation limit 
of $\lg{\frac{L_{\rm x}}{L_{\rm bol}}} \sim -3$ which builds the 
upper envelope for the X-ray emission from TTS.

The M9-type BD LH\,0429+17 was originally listed by 
\citey{Leggett89.1} as a candidate member of the 
Hyades cluster. However, both the presence of Li as an indication of youth 
and proper motion measurements revealed that it belongs to the Taurus
star forming region located in the background of the Hyades 
(\cite{Reid99.1}). 
If LH\,0429+17 were actually associated with the Hyades, and its 
distance was three times smaller than assumed, this would not 
shift the object towards the expected relation in Fig.~\ref{fig:bol}, as
the values for 
$\lg{L_{\rm x}}$ and $\lg{L_{\rm bol}}$ are both affected in the same 
direction. 

Concerning the \sO \ objects we stress that their bolometric luminosity 
is subject to considerable uncertainty.
We based our calculations of the bolometric correction 
on $I$ band magnitudes. 
For spectral type M6 and later, we used values for $BC_{\rm I}$ from 
\citey{Comeron00.1}. For earlier spectral types we used the data for main 
sequence stars given by \citey{Kenyon95.1} which probably come close 
to those of our pre-main sequence objects, but may not be precise.
Further uncertainty derives from the fact
that the spectral types for all three detected objects in \sO \ were 
only estimated from their colors, 
and are uncertain by $\sim$\,2 subclasses. 
As a consequence $BC_{\rm I}$, and hence $L_{\rm bol}$  might 
have been underestimated.  
Finally, the uncertainty in the distance measurement of the \sO \ association is remarkably high. The error of the mean parallax of $\pi = 2.84$\,mas measured by {\em HIPPARCOS} is $\pm 0.91$\,mas, resulting in a distance interval from 267\,pc to 518\,pc. This, however, together with the fact that membership to the cluster is not yet confirmed for all objects of the sample (which would affect their distance) does not result in a change of $\lg{\frac{L_{\rm x}}{L_{\rm bol}}}$ as $L_{\rm x}$ and $L_{\rm bol}$ are affected in a similar way. 

Both for \sO \ and Tau, we can also exclude that the unusually 
high values for $L_{\rm x}$ have their origin in a flare as the 
objects under consideration do not show significant variability 
(see Sect.~\ref{subsect:variability}).
We also checked for indications of circumstellar disks,
which would result in a higher extinction and resulting
mis-estimate of $L_{\rm x}$. (Note that it is unclear whether 
extinction by accretion disks has similar or differing
effects on  $L_{\rm x}$ and $L_{\rm bol}$.) However, $J$, $H$, and $K$ measurements
available from the 2MASS catalogue show no evidence for near infrared
excess pointing at the presence of disks in any of the
three X-ray detected VLM objects in $\sigma$\,Ori.
Similarly, from our recently obtained L-band photometry 
of a sample of $\sigma$\,Ori BDs -- not coincident with 
the X-ray emitting sample -- $K-L$ excess is revealed in only
1 out of 6 examined objects (\cite{Jayawardhana02.1}).
%


In the sample in the ONC quite a few VLM objects show 
significantly higher $\lg{\frac{L_{\rm x}}{L_{\rm bol}}}$ ratios than 
$-3$. Together with the \sO \ association the ONC belongs to the
youngest star forming regions. Therefore, we argue that a
high $\lg{\frac{L_{\rm x}}{L_{\rm bol}}}$ ratio could be a property 
of very young low-mass stars and/or BDs. 

\subsection{Other activity parameters}\label{subsect:activity}

To examine the activity near the substellar boundary in more detail we compared the 
X-ray emission of the young BDs and BD candidates to their 
H$\alpha$ emission and rotation rates.
No rotation periods are known for the objects of our sample. 
The only published measurements of rotation for young BDs so far are the 
$v \sin{i}$ values of six BDs in the Cha\,I complex, given by 
\citey{Joergens01.1}. Those data do not yield any correlation between 
X-ray activity and $v \sin{i}$. 

\begin{figure}
\begin{center}
\resizebox{8.5cm}{!}{\includegraphics{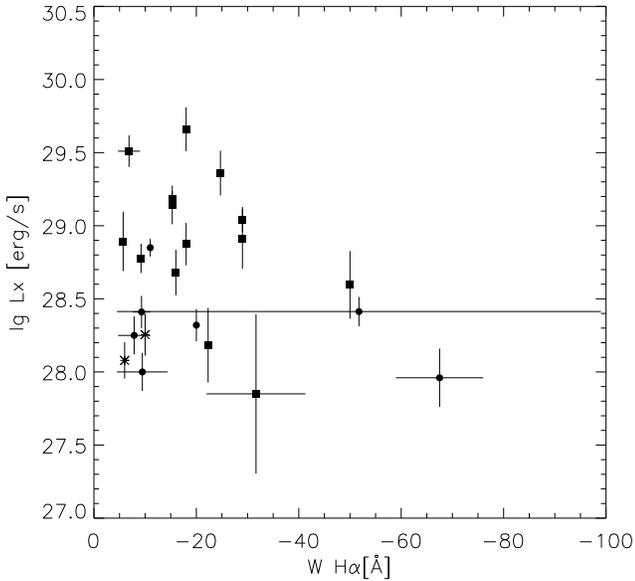}}
\caption{Relation between $\lg {L_{\rm x}}$ and $\rm W_{H_{\alpha}}$ for objects from Fig.~\ref{fig:bol}. 
This sample, however, is smaller, as $\rm W_{H_{\alpha}}$ is not known for all objects. Plotting symbols have the same meaning as above. For $\rm W_{H_{\alpha}}$ measurements see references in Table~\ref{tab:all_bds_xL}.}
\label{fig:alpha}
\end{center}
\end{figure}

The data plotted in Fig.~\ref{fig:alpha} show the relation between 
$\lg {L_{\rm x}}$ and the equivalent width of the H$\alpha$ emission line. 
The vertical bars show the error in $\lg {L_{\rm x}}$, but the horizontal 
bars indicate the range between the lowest and highest value that was 
measured for the H$\alpha$ equivalent width. 
The data suggests that the X-ray luminosity decreases with
increasing H$\alpha$ emission.
However, the equivalent width of H$\alpha$ depends on 
spectral type, and more universal measures of chromospheric
emission such as $L_{\rm H\alpha}$ are clearly required.
The trend shown in Fig.~\ref{fig:alpha} and discussed above reminds of the 
tendency observed for the (higher-mass) TTS in Taurus, 
where classical TTS, i.e. those with accretion disks, show stronger
H$\alpha$ but weaker X-ray emission than the diskless weak-line TTS.
In this sense H$\alpha$ emission would probe accretion rather than
chromospheric activity. 

Most recently, \citey{Natta01.1} reported that for the BD 
ChaH$\alpha$\,1 (the object with strongest H$\alpha$ variability 
in our Fig.~\ref{fig:alpha})   
the observed IR emission is well described by a
circumstellar disk model analogous to that for a TTS. 
For most of the BDs and BD candidates from the sample shown in 
Fig.3 no evidence for disks has been presented yet.
But a number of recent studies have shown that some young substellar
objects have $JHKL$ excesses (\cite{Wilking99.1}, \cite{Muench01.1})
indicative of disks.
More attention to the evolutionary status of these objects is clearly
needed.

\begin{figure}
\begin{center}
\resizebox{8.5cm}{!}{\includegraphics{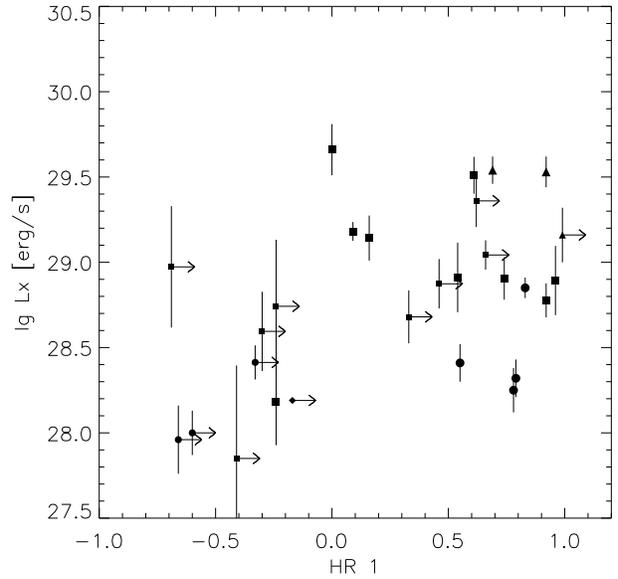}}
\caption{Relation between $\lg{L_{\rm x}}$ and $HR\,1$ for 
all {\em ROSAT} PSPC detected young BDs and BD candidates. Plotting symbols are defined as in the figures above. Arrows denote upper limit measurements.}
\label{fig:hr1}
\end{center}
\end{figure}

The relation between X-ray luminosity and PSPC hardness ratio $HR$\,1 is shown
in Fig.~\ref{fig:hr1}. The arrows denote upper limits in $HR$\,1, i. e. there 
were no source counts in the soft band, so the maximum number of
photons for the S-band was extracted from the background. 
Both the $HR$\,1-ratios and its upper limits indicate that the X-ray 
emission of the {\em ROSAT} detected VLM objects in \sO \ and Tau 
is stronger in the hard band,
a sign of strong magnetic activity and youth. 
A tendency towards larger $HR1$ for
objects with higher $L_{\rm x}$ is observed, which is in line with results by 
\citey{Preibisch97.1} who showed for a sample of T Tauri stars that 
$L_{\rm x}$ rises with increasing coronal temperature.

\begin{figure}
\begin{center}
\resizebox{8.5cm}{!}{\includegraphics{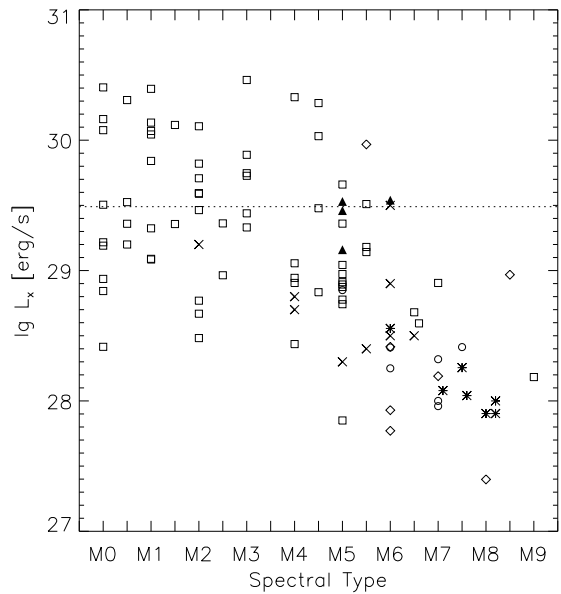}}
\caption{Relation between $\lg {L_{\rm x}}$ and spectral type. Symbols have 
the same meaning as in the figures above. For reasons of clarity, however, 
we left all symbols empty, except for the \sO \ objects. In addition to the BDs and BD candidates discussed 
in this paper, we added the sample of T Tauri stars in the Taurus-Auriga region 
from SN01 for comparison. The dashed line denotes the completeness limit for the {\em ROSAT} PSPC observation of the \sO \ cluster. Note that 
the spectral types of the \sO \ members are only rough values estimated from 
their colors. For some of the objects in the ONC, spectral types are given as an 
interval (see \protect\cite{Feigelson02.1}); in those cases we plot 
the mean value.}
\label{fig:LxSpT}
\end{center}
\end{figure}

\begin{figure}
\begin{center}
\resizebox{8.5cm}{!}{\includegraphics{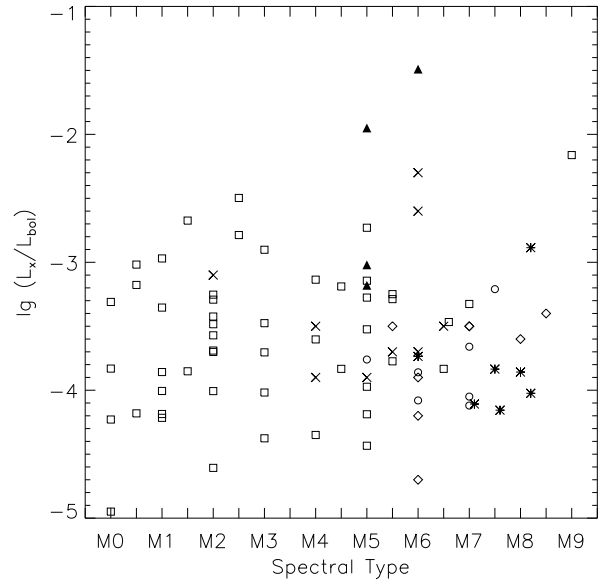}}
\caption{Relation between $\lg {\frac{L_{\rm x}}{L_{\rm bol}}}$ and spectral type. Symbols have the same meaning as in the figures above.
}
\label{fig:LxLbolSpT}
\end{center}
\end{figure}

In Figs.~\ref{fig:LxSpT} and \ref{fig:LxLbolSpT} we plot $\lg{L_{\rm x}}$ and 
$\lg{ \frac{L_{\rm x}}{L_{\rm bol}}}$ as a function of spectral type. In addition to 
the VLM objects treated so far we add the sample of X-ray emitting M-type T 
Tauri stars in the Taurus-Auriga region studied by SN01 from 
spectral type M0 to M4.5. There is a monotonic decrease in $\lg{L_{\rm x}}$ 
towards later spectral types. $\lg{\frac{L_{\rm x}}{L_{\rm bol}}}$, however, which is independent of radius and hence of the emitting area, does not 
change significantly at the transition to substellar masses near 
spectral type $\sim$\,M7. 


\subsection{X-ray Variability}\label{subsect:variability}

Visual inspection of the X-ray lightcurves in the sample of 
\citey{Neuhaeuser99.1} presented no evidence for variability.
Here, we performed a systematic search for variability in all {\em ROSAT}
detected young BDs and BD candidates making use of the 
unbinned Kolmogorov-Smirnov (KS) test. We applied the KS-test on the 
photon arrival times of each source  
after removing data gaps which are due to the satellite operation.
We also checked for variability in the background at a nearby source-free
region using the same approach. The individual results for the objects in $\sigma$\,Orionis and Taurus 
are given in the last column of 
Table~\ref{tab:bds}. 
None of the 6 {\em ROSAT} detected BDs and BD candidates in Cha\,I which are clearly
resolved (\cite{Comeron00.1}) 
turned out to be variable. The only BD candidate in $\rho$\,Oph detected by 
\citey{Neuhaeuser99.1} shows no variability either.
The total number of objects displaying variability 
above the 95\,\% 
level is 4 (out of 23 examined cases). 
\citey{Preibisch02.1} observed variability in only one out of
the 7 {\em Chandra} detected BDs/BD candidates in IC\,348.
For a detailed study of variability of the X-ray emission from BDs and BD candidates there is a need of longterm study including longer, continuous  observing intervals.

\section{Summary and Conclusions}\label{sect:conclusions}

We have searched for X-ray emission from VLM objects near and below the substellar limit in the young stellar associations \sO \ and Tau. Three objects in \sO \ and 13 in Tau were detected in {\em ROSAT} PSPC and HRI observations. We combined these results with all X-ray detections from BDs and BD candidates in star forming regions available up to date and studied stellar activity parameters such as X-ray emission, the ratio $\lg {\frac{L_{\rm x}}{L_{\rm bol}}}$, H$\alpha$ emission and X-ray hardness ratios of the whole sample. The comparison of the X-ray emission of the young BDs and BD candidates with that of (higher-mass) TTS in Taurus shows that $L_{\rm x}$ decreases monotonically into the BD regime. However, $\lg {\frac{L_{\rm x}}{L_{\rm bol}}}$ remains approximately constant suggesting that the efficiency at which hot coronal gas is produced does not change.
For a statistical evaluation of X-ray variability we applied the KS-test
to all {\em ROSAT} detected BDs and BD candidates in star forming regions.
We found that $\sim 17$\,\% 
of the sources are variable with $> 95$\,\% 
confidence. 
To complete the study on activity parameters in young VLM stellar and substellar objects, there is a clear need for rotation and  H$\alpha$ measurements and a search for infrared excess giving clues for the presence of circumstellar disks.

\begin{acknowledgements}
We would like to thank R. Neuh\"auser for constructive discussions and for reading the manuscript.
BS wishes to acknowledge financial support from the BMBF through the DLR 
under grant number 50-OR-0104. 
The {\em ROSAT} project is supported by the Max-Planck-Gesellschaft and
the German federal government (BMBF/DLR).
\end{acknowledgements}


\begin{thebibliography}{}

\bibitem[\protect\astroncite{Barrado y~Navascues et~al.}{2001}]{Barrado01.1}
Barrado y Navascues D., Zapatero Osorio M. R., B\'ejar V. J. S., et al., 2001, A\&A 377, L9

\bibitem[\protect\astroncite{Basri}{2000}]{Basri00.1}
Basri G., 2000, ARAA 38, 485

\bibitem[\protect\astroncite{Basri}{2001}]{Basri01.1}
Basri G., 2001, In: 11th Cambridge Workshop on Cool Stars, Stellar Systems and the Sun, R. J. Garc\'\i a Lopez R, Rebolo \& M. R. Zapaterio Osorio (eds.), A.S.P. Conf. Ser. 223, 261

\bibitem[\protect\astroncite{B\'ejar et~al.}{1999}]{Bejar99.1}
B\'ejar V. J. S., Zapatero Osorio M. R. \& Rebolo R. 1999, ApJ 521, 671

\bibitem[\protect\astroncite{B\'ejar et~al.}{2001}]{Bejar01.1}
B\'ejar V. J. S., Mart\'\i n E. L., Zapatero Osorio M. R., et al. 2001, ApJ 556, 830 

\bibitem[\protect\astroncite{Comer\'on et~al.}{2000}]{Comeron00.1}
Comer\'on F., Neuh\"auser R. \& Kaas A. A., 2000, A\&A 359, 269

\bibitem[\protect\astroncite{Cruddace et~al.}{1987}]{Cruddace87.1}
Cruddace R. G., Hasinger G. R. \& Schmitt J. H., 1987, in: Murtagh F. \& Heck A. (eds.) ESO Conference and Workshop Proc. 28, 177


\bibitem[\protect\astroncite{Elias}{1978}]{Elias78.1}
Elias J. H., 1978, ApJ 224, 857

\bibitem[\protect\astroncite{Feigelson \& Nelson}{1985}]{Feigelson85.1}
Feigelson E. D. \& Nelson P. I., 1985, ApJ 293, 192

\bibitem[\protect\astroncite{Feigelson et~al.}{2002}]{Feigelson02.1}
Feigelson E. D., Bross P., Gaffney III, J. A., et al., 2002, ApJ, in press

\bibitem[\protect\astroncite{Fleming et~al.}{1995}]{Fleming95.1}
Fleming T. A., Schmitt J. H. M. M. \& Giampapa M. S., 1995, ApJ 450, 401

\bibitem[\protect\astroncite{Garmire et~al.}{2000}]{Garmire00.1}
Garmire G., Feigelson E. D., Broos P., et al., 2000, ApJ 120, 1426

\bibitem[\protect\astroncite{Gizis et~al.}{2000}]{Gizis00.1}
Gizis J. E., Monet D. G., Reid I. N., et al., 2000, AJ 120, 1085

\bibitem[\protect\astroncite{Imanishi et~al.}{2001}]{Imanishi01.1}
Imanishi K., Tsujimoto M. \& Koyama K., 2001, ApJ 563, 361

\bibitem[\protect\astroncite{Jayawardhana et~al.}{2002}]{Jayawardhana02.1}
Jayawardhana R., Ardila D. \& Stelzer B., 2002, In: Brown Dwarfs, IAU\,Symp. No. 211, E. L. Martin (ed.), in press

\bibitem[\protect\astroncite{Joergens \& Guenther}{2001}]{Joergens01.1}
Joergens V. \& Guenther E., 2001, A\&A 379, L9

\bibitem[\protect\astroncite{Kenyon \& Hartmann}{1995}]{Kenyon95.1}
Kenyon S. J. \& Hartmann L., 1995, ApJ SS, 101, 117


\bibitem[\protect\astroncite{Lee}{1968}]{Lee68.1}
Lee T. A., 1968, AJ 152, 913

\bibitem[\protect\astroncite{Leggett \& Hawkins}{1989}]{Leggett89.1}
Leggett S. K. \& Hawkins M. R. S., 1989, MNRAS 238, 145

\bibitem[\protect\astroncite{Mart\'\i n~et~al.}{2001a}]{Martin01.1}
Mart\'\i n E. L., Zapatero Osorio M. R., Barrado y Navascues D., et al., 2001a, ApJ 558, L117

\bibitem[\protect\astroncite{Mart\'\i n~et~al.}{2001b}]{Martin01.2}
Mart\'\i n E. L., Dougados C., Magnier E., et al., 2001b, ApJ 561, L195

\bibitem[\protect\astroncite{Mohanty \& Basri}{2002}]{Mohanty02.1}
Mohanty S. \& Basri G., 2002, astro-ph/0201455

\bibitem[\protect\astroncite{Muench et~al.}{2001}]{Muench01.1}
Muench A. A., Alves J., Lada C. J \& Lada E. A., 2001, ApJ 558, L51

\bibitem[\protect\astroncite{Natta \& Testi}{2001}]{Natta01.1}
Natta A. \& Testi L., 2001, A\&A 372, L22

\bibitem[\protect\astroncite{Neuh\"auser et~al.}{1995}]{Neuhaeuser95.1}
Neuh\"auser R., Sterzik M. F., Schmitt J. H. M. M., Wichmann R. \& Krautter J., 1995, A\&A 297, 391

\bibitem[\protect\astroncite{Neuh\"auser \& Comer\'on}{1998}]{Neuhaeuser98.1}
Neuh\"auser R. \& Comer\'on F., 1998, Science 282, 83

\bibitem[\protect\astroncite{Neuh\"auser \& Comer\'on}{1999}]{Neuhaeuser99.2}
Neuh\"auser R. \& Comer\'on F., 1999, A\&A 350, 612

\bibitem[\protect\astroncite{Neuh\"auser et~al.}{1999}]{Neuhaeuser99.1}
Neuh\"auser R., Brice\~no C., Comer\'on F., et al., 1999, A\&A 343, 883

\bibitem[\protect\astroncite{Preibisch}{1997}]{Preibisch97.1}
Preibisch T., 1997, A\&A 320, 525

\bibitem[\protect\astroncite{Preibisch \& Zinnecker}{2001}]{Preibisch01.1}
Preibisch T. \& Zinnecker H., 2001, AJ 122, 866

\bibitem[\protect\astroncite{Preibisch \& Zinnecker}{2002}]{Preibisch02.1}
Preibisch T. \& Zinnecker H., 2002, AJ 123, 1613

\bibitem[\protect\astroncite{Reid \& Hawley}{1999}]{Reid99.1}
Reid I. N. \& Hawley S. L., 1999, AJ 117, 343

\bibitem[\protect\astroncite{Rutledge et~al.}{2000}]{Rutledge00.1}
Rutledge R. E., Basri G., Mart\'\i n E. L. \& Bildsten L., 2000, ApJ 538, L141

\bibitem[\protect\astroncite{Stelzer \& Neuh\"auser}{2001}]{Stelzer01.1}
Stelzer B. \& Neuh\"auser R., 2001, A\&A 377, 538 (SN01)

\bibitem[\protect\astroncite{Walter}{1997}]{Walter97.1}
Walter F. M. 1997, MemSai 68, 1081

\bibitem[\protect\astroncite{Wilking et~al.}{1999}]{Wilking99.1}
Wilking B. A., Greene T. P. \& Meyer M. R., 1999, AJ 117, 469

\bibitem[\protect\astroncite{Zapatero et~al.}{2000}]{Zapatero00.1}
Zapatero Osorio M. R., B\'ejar V. J. S., Mart\'\i n E. L., et al. 2000, Science 290, 103

\bibitem[\protect\astroncite{Zapatero et~al.}{2002}]{Zapatero02.1}
Zapatero Osorio M. R., B\'ejar V. J. S., Pavlenko Ya., et al. 2002, A\& A 384, 937  

\end{thebibliography}
\end{document}